\documentstyle[12pt]{article}
\normalsize

\let\oldtheequation=\theequation
\def\doteqs#1{\setcounter{equation}{0}
            \def\theequation{{#1}.\oldtheequation}}

\newcounter{sxn}
\def\sx#1{\addtocounter{sxn}{1} \bigskip\medskip \goodbreak \noindent{\large\bf
\centerline{\thesxn.~~#1}} \nobreak \medskip}
\def
\sxn#1{\sx{#1} \doteqs{\thesxn}}

\newcounter{axn}

\def\br{\begin{thebibliography}{abc}}
\def\er{\end{thebibliography}}
\def\rf{\bibitem}

\date{}

\begin{document}

\bibliographystyle{unsrt}
\footskip 1.0cm
\thispagestyle{empty}
\setcounter{page}{0}
\begin{flushright}
SU-4240-507\\
May 1992
\end{flushright}
\vspace*{10mm}

\begin{center}
{\LARGE THE EDGE STATES OF THE {\it BF} SYSTEM \\
               AND \\
   THE LONDON EQUATIONS \\}
\vspace*{6mm}
{\large A.P. Balachandran and P. Teotonio-Sobrinho}\\
\vspace*{5mm}
{\it Physics Department, Syracuse University,\\
 Syracuse, NY, 13244-1130, USA\\}
\end{center}
\vspace*{2cm}

\normalsize
\centerline{\bf Abstract}
\vspace*{5mm}

It is  known that the 3d Chern-Simons interaction describes the scaling limit
of a quantum Hall system and predicts edge currents in a sample with boundary,
the currents generating a chiral $U(1)$ Kac-Moody algebra. It is no doubt also
recognized that in a somewhat similar way, the 4d $BF$ interaction (with $B$ a
two form, $dB$ the dual $^*j$ of the eletromagnetic current, and F the
electromagnetic field form) describes the scaling limit of a superconductor. We
show in this paper that there are edge excitations in this model as well for
manifolds with boundaries. They are the modes of a scalar field with invariance
under the group of diffeomorphisms (diffeos) of the bounding spatial
two-manifold. Not all of this group seem implementable by operators in quantum
theory, the implementable group being a subgroup of volume preserving diffeos.
The $BF$ system in this manner can lead to the $w_{1+\infty }$ algebra and its
variants. Lagrangians for fields on the bounding manifold which account for the
edge observables on quantization are also presented. They are the analogues of
the $1+1$ dimentional massless scalar field Lagrangian describing the edge
modes of an abelian Chern-Simons theory with a disk as the spatial manifold. We
argue that the addition of ``Maxwell'' terms constructed from $F\wedge ^*F$ and
$dB\wedge ^*dB$ do not affect the edge states, and that the augmented
Lagrangian has an infinite number of conserved charges- the aforementioned
scalar field modes- localized at the edges. This Lagrangian is known to
describe London equations and a massive vector field. A $(3+1)$ dimensional
generalization of the Hall effect involving vortices coupled to $B$ is also
proposed.

\newpage
\newcommand{\be}{\begin{equation}}
\newcommand{\ee}{\end{equation}}
\baselineskip=24pt

\sxn{INTRODUCTION}\label{sec-introduction}

When a physical system undergoes spontaneous symmetry breakdown, its
behavior at low energy and momentum is well approximated by the dynamics of a
Nambu-Goldstone mode. The latter is a field valued in a homogeneous space
$G/H$, $G$ being the Lagrangian symmetry group and $H$ the unbroken one.

The group $G$ is a gauged symmetry group for numerous physical systems. That is
the case in electroweak theory\cite{eletroweak} which involves the
spontaneous reduction of $SU(2)\times U(1)$ to the $U(1)$ group of
electromagnetism. It is the case in superconductivity\cite{super} where the
electromagnetic $U(1)$ breaks down to the discrete group $Z_2$.

Many years ago, it was pointed out by London\cite{super} that the essential
phenomenology of superconductivity is captured by the constituent equation
\be
\partial _\mu J_\nu -\partial _\nu J_\mu =\lambda F_{\mu \nu
},\,\,\,\,\,\,\lambda ={\rm constant}
\label{eq:1.1}
\ee
relating
the current $J_\mu $ to the electromagnetic field $F_{\mu \nu }$. The approach
to superconductivity based on the Nambu-Goldstone mode incorporates this
fundamental equation. Thus for superconductivity, the mode is a complex field
$e^{i\varphi }$ of unit modulus and charge $2e$ and responds to the gauge
transformation $A_\mu \rightarrow A_\mu +\frac 1e\partial _\mu \Lambda $
according to $e^{i\varphi }\rightarrow e^{2i\Lambda }e^{i\varphi }$. If
$\langle H\rangle $ is the (real) vacuum value of the Higgs or order
parameter field $H$, then the current
$$
J_\mu =4ie\langle H\rangle ^2e^{-i\varphi }D_\mu e^{i\varphi }\,\, ,
$$
\be
D_\mu =\partial _\mu -2ieA_\mu \label{eq:1.2}
\ee
of the Landau-Ginsburg Lagrangian\cite{super}  in the London limit is gauge
invariant. Furthermore, the London ansatz~(\ref{eq:1.1}) is an identity with
$\lambda =8e^2\langle H\rangle ^2$.

There is an alternative approach to the London ansatz which leads to the $BF$
system
and which is of particular interest in this paper. It begins with the remark
that $J_\mu $ fulfills the continuity equation
\be
\partial ^\mu J_\mu =0 \label{eq:1.3}
\ee
in addition to~(\ref{eq:1.1}). (Our metric has signature $-,+,+,+$.) Its
expression in~(\ref{eq:1.2}) based on the Nambu-Goldstone field can be thought
of as solving~(\ref{eq:1.1}) as an identity and obtaining~(\ref{eq:1.3}) as a
field equation from the Lagrangian
\be
\int d^3x\left\{ -\langle H\rangle ^2\left( D^\mu e^{i\varphi }\right)
^{*}\left( D_\mu e^{i\varphi }\right) -\frac 14F^{\mu \nu }F_{\mu \nu }\right\}
.\label{eq:1.4}
\ee

In the alternative approach, we solve the continuity equation instead as an
identity by setting
\be
J_\mu =-\epsilon _{\mu \nu \lambda \rho }\partial ^\nu B^{\lambda \rho }
\label{eq:1.5}
\ee
where the convention $\epsilon ^{0123}=+1$ is adopted for the Levi-Civita
symbol $\epsilon ^{\mu \nu \lambda \rho }$.
The constituent equation is then obtained as a field equation from the
Lagrangian
$$L=\int d^3x{\cal L},$$
$$
{\cal L}=\frac 12\epsilon ^{\mu \nu \lambda \rho }B_{\mu \nu }F_{\lambda \rho }
-\frac {1}{3\lambda }H^{\mu \nu \lambda }H_{\mu \nu \lambda }
-\frac 14F^{\mu \nu }F_{\mu \nu }\,\, ,
$$
\be
H_{\mu \nu \lambda }=\partial _\mu B_{\nu \lambda }+\partial _\nu
B_{\lambda
\mu }+\partial _\lambda B_{\mu \nu }\,\, .
\label{eq:1.6}
\ee

Lagrangians of this sort have come into prominence in modern times in
connection with topological field theories\cite{witten,horwitz} and exotic
statistics\cite{anezires,liu}, as a method to generate mass for gauge
fields distinct from the Higgs mechanism\cite{bowick} and in connection with
quantum hair for black holes \cite{lahiri}.
For spatial manifolds
devoid of boundaries, the classical and quantum aspects of ${\cal L}$
have been developed to an advanced level
particularly
by Allen et al.\cite{bowick}. In this paper, we propose to investigate
(\ref{eq:1.6})  when the underlying manifold $\Sigma $ has a boundary. It can
for instance be a three dimensional ball, a solid torus, or a solid cylinder.
For reasons of simplicity, we will exclusively consider these examples and
assume
also that $\Sigma$ is a submanifold of $\Re ^3$ (with the metric induced from
the Euclidean metric of $\Re ^3$ and with
local Cartesian coordinates $x^i$). We will show that there are edge states
localized at
the boundary of $\Sigma $ which are similar to the edge states in the quantum
Hall system[8-11]. There is for example a diffeomorphism
(diffeo) group (or a central extension thereof) acting on them just as the
Virasoro group\cite{goddard} acts on the Hall edge
states\cite{witten,bbgs1,ajit}.

Our central attention in this paper is focussed on these edge states. Their
basic features are not sensitive to the dynamics in the interior $\Sigma ^0$ of
$\Sigma $. It is therefore possible, although not necessary, to assume the
minimum energy configuration in $\Sigma ^0$ with no essential loss of content.
With this assumption, the edge states are described by the $BF$
system \cite{horwitz,anezires,witten} with the Lagrangian
$$L^{*}=\int d^3x{\cal L}^{*}\,\, ,$$
\be
{\cal L}^{*}=\frac 12\epsilon ^{\mu \nu \lambda \rho }B_{\mu \nu }F_{\lambda
\rho }\,\, .\label{eq:1.7}
\ee
In this way, we discover that the edge states are well accounted for by a
topological field theory.

In Section~2, we  briefly describe the canonical formalism for
 ~(\ref{eq:1.6}) and its edge observables.  The important result that these
observables form an infinite number of constants of motion, with a relation to
gauge transformations similar to that of charges in gauge theories, is also
established. The existence of such an infinite number of constants of motion
(or
edge observables) depends only on gauge invariance and the presence of the
boundary $\partial \Sigma $, and not on the specific Lagrangian~(\ref{eq:1.6}).
We also display the canonical generators of the aforementioned diffeo group and
establish their Sugawara form\cite{goddard}. Next, in Section~3, we pull back
or restrict
 the canonical formalism to the part of the phase space with zero energy in
$\Sigma ^0$. It results in the $BF$ Lagragian (\ref{eq:1.7}). All these
activities are as yet classical.

Section~4 quantizes the basic edge observables of Section 2
for $\Sigma $ a ball or a solid torus. (The solid cylinder
has a special interest, having an association with the $w_{1+\infty }$
algebra\cite{pope}. It is hence separately discussed in Section 5,
although much of what is did here applies there as well.)
They are described by an infinite number of independent harmonic oscillators
which can be associated with the modes of a scalar field. There is also a
Lagrangian defined on $\partial \Sigma $ which leads to these excitations,
namely
\be
L^2\int _{\partial \Sigma }\partial _t\Psi dC,\,\,\,\,\,L={\rm constant}
\label{eq:1.8-i}
\ee
where $\Psi $ is a scalar field and $C$ a one form. It is the precise analogue
of the Lagrangian
\be
\pm L^2\int _{S^1}\partial _t\varphi \, \partial _\theta \varphi \, d\theta
\label{eq:1.9-i}
\ee
which describe the edge observables for the Chern-Simons field on a disc $D$
with the circle $S^1$ as its boundary $\partial D$.

Now there is much to be said about (\ref{eq:1.8-i}). It has for instance to be
quantised. There is also a second order Lagrangian which leads to
(\ref{eq:1.8-i}) just as the Lagrangian
\be
L^3_1 \int \left[ (\partial _t \varphi )^2 -(\frac {1}{L_2}\partial _\theta
\varphi )^2\right]d\theta ,\,\,\, L_i={\rm constants}\label{eq:1.10-i}
\ee
leads to (\ref{eq:1.9-i}) on chiral decomposition. We will discuss these
matters elsewhere\cite{we}. In this paper, we will instead discuss the
diffeo group acting on the edge states in a little detail. We have alluded to
this group before, while its existence is also suggested by the invariance of
(\ref{eq:1.8-i}) under the diffeos of $\partial \Sigma $.

The following point is worthy of note in this context. The association of edge
observables with a quantum scalar field is not unique. This is because the
definition of the latter involves the choice of a dispersion relation
connecting frequency and mode labels. It is with its help that we normally
define creation and annihilation operators and a Fock space. [~A~more general
method of quantisation involving the choice of a complex structure and a
metric on the phase space of fields\cite{rajeev} is also possible. For our
present purposes, it is enough to consider the method based on dispersion
relation.] This dispersion relation is not given by (\ref{eq:1.6}),
(\ref{eq:1.7}) or (\ref{eq:1.8-i}) and must be supplied externally. [But the
second order Lagrangian analogous to (\ref{eq:1.10-i}) does provide adequate
data for quantisation, as we
shall discuss in \cite{chiral}.] An analogous
condition prevails in the Chern-Simons problem\cite{bbgs1,bbgs2}, where too
the
quantum edge field can be shown to be indeterminate without the dispersion
relation
as a new input. The latter can be constructed (although still not uniquely) by
the insistence that the
Virasoro algebra acts properly on the Fock space of the scalar field.

But a similar
approach does not quite work here, for (\ref{eq:1.6}), (\ref{eq:1.7}) or
(\ref{eq:1.8-i}). This is because, in so far as we can tell, it is impossible
to implement the Lie algebra of the entire $\partial \Sigma $ diffeo group (or
a possible central extension thereof) on a Fock space.( We will not be too
careful in the Introduction to distinguish a diffeo group from its central
extension.) We argue that the implementable
diffeos are those preserving a volume form $\mu $ on $\partial \Sigma $ ( at
least for simple choices $\partial \Sigma $). This group of diffeos is often
denoted by
$SDiff(\partial \Sigma )$ (and its identity component by $SDiff_0(\partial
\Sigma )$) in the
literature\cite{sdiff}. It preserves the Lagrangian
\be
L=\frac 12\int _{\partial \Sigma }\mu \left( (\partial _t \varphi )^2-\omega
_0^2\varphi ^2\right) \label{eq:1.8}
\ee
of the field $\varphi$, $\omega _0$ being a positive constant.
The canonical treatment of (\ref{eq:1.8}) leads to essentially all our edge
states and implementable
infinitesimal diffeos. Thus, (\ref{eq:1.8}) also may provide a
satisfactory account of our edge states.
The constant $\omega _0$ in (\ref{eq:1.8}) can not be determined by any
reasonable
consideration based on (\ref{eq:1.6}) or (\ref{eq:1.7}), in the same way
that the speed of the field in (\ref{eq:1.10-i}) can not be inferred from the
Chern-Simons action.

The Lagrangian (\ref{eq:1.8}) is the two dimensional analogue of the Einstein
model for specific heat in one dimension\cite{esp.heat}.

Section~5 deals with the solid cylinder for which $\partial
\Sigma =S^1\times \Re ^1$. In this case, with $\mu $ a rotationally and
translationally invariant volume form, $SDiff_0(\partial \Sigma )$ is the
group with the algebra
$w_{1+\infty }$ of conformal field theories\cite{pope}. The group which occurs
in quantum theory is perhaps not this one, but a central extension thereof, but
we have not done the necessary calculations to verify this possibility.

The fields $A$ and $B$ naturally admit two kinds of sources, namely point
charges and vortices or strings. Elsewhere, the $BF$ Lagrangian with such
sources
has been studied~~in investigations of exotic statistics\cite{anezires,liu}.
In a paper under completion\cite{we}, we will demonstrate that like the
Chern-Simons sources, these sources also get `framed', or rather acquire spin
degrees of freedom, on regularization. Vertex operators for the creation of
these sources,
similar to the Fubini-Veneziano vertex operator \cite{goddard}, will be
constructed and a
spin-statistics theorem for vortices connecting a certain exchange and a
$2\pi $
rotation of the spin variables alluded to above will be established. It must
be
remarked that neither this exchange nor this rotation are those appropriate
for
the conventional spin-statistics theorem so that the result shown here is
genuinely novel. As for the conventional theorem, that too can be proved, in
fact easily and without recourse to relativistic quantum fields, as will be
seen in that paper.

The quantum Hall edge states have a simple physical interpretation\cite{ajit}.
There is an analogue of the Hall effect for vortices coupled to $B$ and we
shall
explain it in Section~6. It is our expectation that the edge
states of (\ref{eq:1.6}) or (\ref{eq:1.7}) will find a similar interpretation
in the context of this phenomenon. Such an interpretation will be helpful for
the observation of these states and merits attention.

A basic physical issue not addressed in this paper and \cite{we} concerns the
reproduciblility of their results in the Higgs field description. Its answer
seems affirmative. We plan to report on this matter, and on the nonabelian
version of the preceding edge states and sources, in future publications.

In references \cite{bbgs1,bbgs2}, the edge states of the Chern-Simons
Lagrangian were considered in the absence of the ``Maxwell'' or ``kinetic
energy'' term
for the $U(1)$ Chern-Simons potential $A$ [the analogue of the term in $L$
from the last term in ${\cal L}$ of (\ref{eq:1.6})] and of the similar term
for its nonabelian
counterpart. It is possible however to generalize that work by including these
terms and proceeding along the lines of this paper. This task has been carried
out by the authors of \cite{bbgs1,bbgs2} in unpublished work.

\newpage
\sxn{THE CANONICAL FORMALISM}\label{sec-can.form}

We follow Dirac\cite{dirac} for the canonical treatment of (\ref{eq:1.6}).

Let $\pi _\mu $ and $P_{\mu \nu }$ be the momenta conjugate to $A_\mu $ and
$B_{\mu \nu }$. The Legendre map of (\ref{eq:1.6}) gives the primary
constraints
\be
\pi _0\approx 0,\,\,\,\,\,\,\,\,\,\,\,\,\,\,\,\,\,\,\,\,\,P_{0i}\approx 0,
\label{eq:2.1}\ee
and the expressions
$$\pi _i=F_{oi}+ \epsilon _{ijk}B_{jk},$$
\be
P_{ij}=\frac 4\lambda H_{0ij} \label{eq:2.2}
\ee
where
$$\epsilon _{ijk}=\epsilon ^{0ijk}, \,\,\,\,\,1\leq i,j\leq 3\,,$$
\be
H_{\mu \nu \lambda }=\partial _\mu B_{\nu \lambda }+\partial _\nu
B_{\lambda
\mu }+\partial _\lambda B_{\mu \nu }
\label{eq:2.3}\ee
and $\approx $ stands for weak equality.

The Hamiltonian is
\begin{eqnarray}
H & = &\int d^3x\left[ \frac 12\left[\pi _i -\epsilon _{ijk}B_{jk}\right]^2+
     \frac {\lambda }{16}P^2_{ij}
     + \frac 14F_{ij}^2 + \frac {1}{3\lambda }H_{ijk}^2 \right. \nonumber \\
  &   &  - A_0\partial _i\pi _i  -B_{0i}(\partial _jP_{ji}
       + 2\epsilon _{ijk}\partial _jA_k)\nonumber \\
  &   &\left. +\psi ^0\pi _0+\psi ^iP_{0i}\right] ,
\label{eq:2.4}
\end{eqnarray}
$\psi ^0$ and $\psi ^i$ being Lagrange multiplier fields. It leads to the
secondary constraints
\be
\partial _i\pi _i\approx 0 ,
\label{eq:2.5}\ee
\be
\partial _jP_{ji}+2\epsilon _{ijk}\partial _jA_k\approx 0.
\label{eq:2.6}\ee
There are no tertiary constraints.

The constraints are all first class. Of these, (\ref{eq:2.1}) eliminate $A_0$
and $B_{0i}$ as observables. They will henceforth be ignored along with $\pi
_0$  and $P_{0i}$, as is permissible.

The Gauss law constraints (\ref{eq:2.5}) and (\ref{eq:2.6}) require delicate
treatment on manifolds with boundary for reasons outlined in \cite{bbgs1}.
Following that work, we first rewrite them in the form
\be
 {\cal G}_0(\lambda ^{(0)})=\int _\Sigma d^3x\lambda ^{(0)} \partial _i\pi
 \approx 0, \label{eq:2.6-a}
\ee
\be
{\cal G}_1(\lambda ^{(1)})=\int _\Sigma d^3x\lambda _i^{(1)} \left[
\partial _jP_{ji}+2\epsilon _{ijk}\partial _jA_k\right] \approx 0 .
\label{eq:2.7}
\ee
The allowed class of ``test functions'' $\lambda ^{(0)}$ and $\lambda ^{(1)}_i$
are then fixed by requiring that
(\ref{eq:2.6-a}) and (\ref{eq:2.7}) are differentiable in the fields $\pi _i$,
$P_{lk}$ and $A_m$. Thus, consider the
variations
\begin{eqnarray}
\delta {\cal G}_0(\lambda ^{(0)})&=& \int _\Sigma d^3x\lambda ^{(0)}
\partial _i\delta \pi _i\nonumber \\ & = &
\int _{\partial \Sigma }d^2x\lambda
^{(0)}n^i \delta \pi _i -\int _\Sigma d^3x\partial _i\lambda ^{(0)}\delta
\pi _i,\label{eq:2.8}
\end{eqnarray}
\begin{eqnarray}
\delta {\cal G}_1(\lambda ^{(1)})&=& \int _\Sigma d^3x\lambda ^{(1)}_i\left[
\partial _j\delta P_{ji}+2\epsilon _{ijk}\partial _j\delta A_k\right]
\nonumber \\ & = &
\int _{\partial \Sigma }d^2x\lambda _i^{(1)}n^j
\left[ \delta P_{ji}+2\epsilon _{ijk}\delta A_k\right]-
\nonumber \\ & &
-\int _\Sigma d^3x\partial _j\lambda _i^{(1)}\left[ \delta
P_{ji} + 2\epsilon _{ijk}\delta A_k\right] ,\label{eq:2.9}
\end{eqnarray}
$n^j$ defining the outward drawn normal to $\partial \Sigma $. The functionals
${\cal G}_j$ are differentiable only if the boundary terms (first terms) in
(\ref{eq:2.8}) and (\ref{eq:2.9}) are absent. In this way, we are led to the
conditions
\be
\lambda ^{(0)}\left| _{\partial \Sigma }\right. =0\,,\,\,\,\,\, \vec n\times
\vec \lambda ^{(1)}\left| _{\partial \Sigma }\right. =0\,.
\label{eq:2.10}
\ee

It is better to write  (\ref{eq:2.6-a}) and (\ref{eq:2.7}) entirely in terms of
forms. Thus let
$${\cal B}\equiv {\cal B}_{ij}dx^idx^j=\frac 12\epsilon _{ijk}\pi _k
dx^idx^j,$$
$${\cal A}\equiv {\cal A}_idx^i=\left[ A_i-\frac 14\epsilon _{ijk}P_{jk}
\right] dx^i,$$
\be
\lambda ^{(1)}=\lambda _i^{(1)}dx^i. \label{eq:2.11}
\ee
(The wedge symbols between differential forms are being omitted.). Then
$${\cal G}_0(\lambda ^{(0)})=\int _\Sigma \lambda ^{(0)}d{\cal B}
\approx 0,$$
\be
{\cal G}_1(\lambda ^{(1)})=2\int _\Sigma \lambda ^{(1)}d{\cal A}
\approx 0\label{eq:2.12}
\ee
if $\lambda ^{(0)},\lambda ^{(1)}$ belong to the test function space ${\cal
I}^{(0)},{\cal I}^{(1)}$, defined by the conditions
\be
\lambda ^{(0)}\left| _{\partial \Sigma }\right.
=0,\,\,\,\,\,\,\,\,\,\,\,\,\,\,\,\,\,\,\,\,\,\,\,\,\,\lambda ^{(1)}\left|
_{\partial \Sigma }\right. =0.\label{eq:2.13}
\ee
Here by the notation $\lambda ^{(N)}\left| _{\partial \Sigma }\right. =0$ for
an $N$ form $\lambda ^{(N)}$, we mean the pull back of $\lambda ^{(N)}$ to $
\partial \Sigma $.

Next consider
$$q(w^{(1)})=\int _\Sigma w^{(1)}{\cal B}\,,$$
\be
p(w^{(2)})=-\int _\Sigma w^{(2)}{\cal A}\label{eq:2.14}
\ee
where $w^{(j)}$ are closed $j$ forms:
\be
dw^{(j)}=0 .\label{eq:2.15}
\ee
(\ref{eq:2.14}) and (\ref{eq:2.15}) are differentiable in $\pi _k$, $P_{jk}$
and $A_l$ even if
\be
w^{(j)}|_{\partial \Sigma }\neq 0 .\label{eq:2.16}
\ee
Their Poisson brackets (PB's) with the constraints vanish.(All PB's are at
equal times.) For example, on using
\be
\left\{ {\cal B}_{ij}(x),{\cal A}_k(y)\right\} =-\frac 12\epsilon _{ijk}\delta
^3(x-y),\label{eq:2.17}
\ee
we get
\begin{eqnarray}
\left\{ p(w^{(2)}),{\cal G}_0(\lambda ^{(0)})\right\} & = &
\int _\Sigma w^{(2)}d\lambda ^{(0)}\nonumber \\
        & = & \int _{\partial \Sigma }w^{(2)}\lambda ^{(0)}=0
\label{eq:2.18}
\end{eqnarray}
by (\ref{eq:2.13}). Thus (\ref{eq:2.14}) are observables.

They are furthermore constants of motion for the Hamiltonian (\ref{eq:2.4}).
This can be shown using the PB's

\begin{eqnarray}
\left\{ q(w^{(1)}),A_i    \right\} & = & -w^{(1)}_i,\nonumber \\
\left\{ q(w^{(1)}),\pi _i \right\} & = & 0,\nonumber \\
\left\{ q(w^{(1)}),B_{ij} \right\} & = & 0,\nonumber \\
\left\{ q(w^{(1)}),P _{ij}\right\} & = & 0 ,\nonumber \\
\left\{ p(w^{(2)}),A_i    \right\} & = &  0,\nonumber \\
\left\{ p(w^{(2)}),\pi _i \right\} & = & -\epsilon _{ijk}w^{(2)}_{jk},
                                          \nonumber \\
\left\{ p(w^{(2)}),B_{ij} \right\} & = & - w^{(2)}_{ij},\nonumber \\
\left\{ p(w^{(2)}),P_{ij} \right\} & = & 0 \label{eq:2.19}
\end{eqnarray}
where $w^{(1)}=w^{(1)}_idx^i$ and $w^{(2)}=w^{(2)}_{jk}dx^jdx^k$ with an
antisymmetric $w^{(2)}_{jk}$. Thus our
dynamical system has an
infinite number of constants of motion. They are the exact analogues of charges
in conventional gauge theories, with a relation to Gauss laws similar to those
of charges \cite{bbgs1}.

Note next that since $q(w^{(1)}+d\lambda ^{(0)})\approx q(w
^{(1)}),\,p(w
^{(2)}+d\lambda ^{(1)})\approx p(w^{(2)})$ in view of (\ref{eq:2.12}) and
(\ref{eq:2.13}), test functions differing by $d\lambda ^{(j-1)}$ define
equivalent (or weakly equal) observables.

Actually, if a $\lambda ^{(j-1)}$ modulo a closed form belongs to ${\cal
I}^{(j-1)}$, then $w^{(j)}$ and $w^{(j)} + d\lambda ^{(j-1)}$
define equivalent observables. This is because a closed form drops out
of $d\lambda ^{(j-1)}$.

When $\Sigma $ is a ball $ B_3$, a closed
$w^{(j)}$ is also exact. In this case, $w^{(j)}=d\xi ^{(j-1)}$ with
$\xi ^{(j-1)}$ and $\xi ^{(j-1)} + \lambda ^{(j-1)}$ giving equivalent
observables. An observable is consequently sensitive only to $\xi
^{(j-1)}|_{\partial \Sigma }$ and can be regarded as localized at the edge.
For suitable choices of $\xi ^{(j-1)}$, they are also localized on an
arbitrary contractible open set at the edge.

The following PB's give the fundamental classical algebra of these edge
excitations:
\begin{eqnarray}
\left\{ q(d\xi ^{(0)}),q(d\bar \xi ^{(0)})\right\} & = &\left\{ p(d\xi
^{(1)}),p(d\bar \xi ^{(1)})\right\} =0, \nonumber \\
\left\{ q(d\xi ^{(0)}),p(d\xi ^{(1)})\right\} & = & \int _\Sigma d\xi
^{(0)}d\xi ^{(1)} \nonumber \\
    & = &
\int _{\partial \Sigma }\xi ^{(0)}d\xi ^{(1)}.\label{eq:2.20}
\end{eqnarray}

Next consider the solid torus ${\bf T}_3$. For ${\bf T}_3$, all $w^{(2)}$ are
exact, $w^{(2)}=d\xi ^{(1)}$. As for $w^{(1)}$, consider $\partial
{\bf T}_3=$ the two-torus $T^2$. It has coordinates $\theta ^1$, $\theta ^2$
with $\theta ^i$ and $\theta ^i + 2\pi $ being identified. Let the cycle
obtained by
increasing $\theta ^2$ by $2\pi $ (with fixed $\theta ^1$) be the one
contractible to a point by shrinking it within ${\bf T}_3$. Then $d\theta ^1$
can be extended from $\partial {\bf T}_3$ to ${\bf T}_3$ as
a globally defined closed but inexact one form on ${\bf T}_3$. Any closed
one form on ${\bf T}_3$ is a linear combination of such an extension and an
exact one form $d\xi ^{(0)}$. Just as for the ball $B_3$, the observables
$q(d\xi ^{(0)})$ and $p(d\xi ^{(1)})$ are localized on $\partial \Sigma $ and
fulfill (\ref{eq:2.20}). For suitable choices of $\xi ^{(j-1)}$, they are also
localized on an arbitrary contractible open set on $\partial \Sigma $.
As for $q(d\theta ^1)$, it too gives an observable living only on $\partial
\Sigma $ since $q(d\theta ^1)\approx q(d\theta ^1 +d\lambda ^{(0)})$. But
$q(\theta ^1)$, or $q(d\theta ^1 +d\xi ^{(0)})$ for any choice of
$\xi ^{(0)}$, are not localized on a contractible open set on $\partial
\Sigma $, $d\theta ^1 +d\xi ^{(0)}$ being cohomologically nontrivial.
$q(d\theta ^1)$ has zero PB with $q(d\xi ^{(1)})$ while
\be
\left\{ q(d\theta ^1),p(d\xi ^{(1)})\right\} = \int _{\partial \Sigma }
\xi ^{(1)}d\theta ^1. \label{eq:2.22-iii}
\ee

The generators of diffeos of $\partial \Sigma $ will now be examined. Let
$\eta =\eta ^i\partial _i$ be a vector field on $\Sigma $ tangent to $\partial
\Sigma $ at $\partial \Sigma $. It generates a diffeo $\Sigma \rightarrow
\Sigma $ and acts on fields by the Lie derivative ${\cal L}_\eta $. For
example,
     $${\cal L}_\eta (\xi ^{(0)})=\eta ^i\partial _i\xi ^{(0)},$$
     $${\cal L}_\eta (\xi ^{(1)})=\left[\partial _i(\xi^{(1)} _j\eta ^j)+
\eta ^j(\partial _j\xi _i^{(1)}-\partial _i\xi _j^{(1)}) \right]dx^i,$$
\be
{\cal L}_\eta \bar \eta ^j\partial _j=\left[ \eta ^i\partial _i,\bar \eta
^j\partial
_j\right] .\label{eq:2.21}
\ee

We want to construct an observable $l(\eta )$ depending on $\eta $ with the
following properties:

1) Its canonical action via PB's is that of an infinitesimal diffeo on the edge
observables:

$$\left\{ l(\eta ),q(w^{(1)})\right\} =q\left( {\cal L}_\eta w
^{(1)}\right),$$
\be
\left\{ l(\eta ),p(w^{(2)})\right\} =p\left( {\cal L}_\eta w
^{(2)}\right)
.\label{eq:2.22}\ee

2) It becomes a constraint if the restriction $\eta |_{\partial \Sigma }$ of
$\eta $ to $\partial \Sigma $ is zero. With this
requirement, it becomes also an edge observable, which is satisfactory as we
are after a theory of edge observables insensitive to $\Sigma ^0$.

It is remarkable that such an $l$ exists. It is
\be
l(\eta )=\int ({\cal L}_\eta {\cal A}){\cal B}.\label{eq:2.23}
\ee

For showing that $l(\eta )$ has the correct properties, the following
identities are useful: Let  $i_{\eta }$ be the contraction on the vector field
$\eta $ so that for example $i_{\eta }A_jdx^j=\eta ^jA_j.$ Then we have
the identity \cite{choquet}
\be
{\cal L}_\eta = di_{\eta }+i_{\eta }d \label{eq:cartan}
\ee
Therefore, if $w=w_{ijk}dx^idx^jdx^k $
is a three form, $w_{ijk}$ being totaly antisymmetric,
\begin{eqnarray}
\int _\Sigma {\cal L}_\eta w & =& \int _\Sigma (di_{\eta }+i_{\eta }
d)w                                               \nonumber \\
                       & = & \int _{\partial \Sigma }i_{\eta }w \nonumber
\end{eqnarray}
by Stokes theorem as $dw =0$. Now $i_{\eta }w $ in the last integral
is $\eta ^iw
_{ijk}dx^jdx^k$ , the differentials being tangent to $\partial \Sigma $. Since
$\eta |_{\partial \Sigma }$ is also similarly tangent, we have $i_{\eta }
w|_{\partial \Sigma }=0$.The basic result
\be
\int _\Sigma {\cal L}_\eta w =0 \label{eq:2.24}
\ee
and its consequence
\be
\int _\Sigma \left( {\cal L}_\eta \alpha \right) \beta =-\int \alpha {\cal
L}_\eta \beta
,\;\;\;\;\alpha =\alpha _idx^i,\,\beta =\beta _{ij}dx^idx^j \label{eq:2.25}
\ee
thus follow.

(\ref{eq:2.25}) helps us to show the differentiability of $l(\eta )$
in all the dynamical fields. Thus if ${\cal A}$ for instance is varied,
\be \delta l(\eta )=\int _\Sigma \left( {\cal L}_\eta \delta {\cal A}\right)
{\cal B}=-\int _\Sigma \delta {\cal A}{\cal L}_\eta {\cal B},
\label{eq:2.27}\ee
which shows its differentiability in ${\cal A}$.

We must now check that $l(\eta )$ is an observable having properties 1) and 2).
It is an observable if it is weakly invariant under the gauge transformations
generated by ${\cal G}_i$. The best way to verify this may be as follows. A
transformation due to ${\cal G}_0(\lambda ^{(0)})$ for instance converts $A$ to
$A + d\lambda ^{(0)}$. Using ${\cal L}_\eta d=d{\cal L}_\eta $\cite{choquet},
we therefore have
\begin{eqnarray}
\left\{ {\cal G}_0(\lambda ^{(0)}),l(\eta )\right\} & = & \int _\Sigma
d({\cal L}_\eta \lambda ^{(0)}){\cal B}\nonumber \\
   & = & -\int _\Sigma {\cal L}_\eta \lambda ^{(0)}d{\cal B}\approx 0
\label{eq:2.28}
\end{eqnarray}
as  ${\cal L}_\eta \lambda ^{(0)}|_{\partial \Sigma }=0$. In a similar way, we
can show its weak invariance under $B\rightarrow B+d\lambda ^{(1)},\,\,\lambda
^{(1)}\in {\cal I}^{(1)}$.

As for 1), it follows easily from (\ref{eq:2.17}). Only 2) now remains. Let
$\bar \eta $ be a vector field vanishing at $\partial \Sigma $. Then
\begin{eqnarray}
l(\bar \eta ) &=&\int _\Sigma \left[ (di_{\bar \eta }+i_{\bar \eta }d){\cal
                              A}\right] {\cal B}\nonumber \\
                 &=&-\int _{\Sigma } (i_{\bar \eta}{\cal A})d{\cal B} + \int
_{\Sigma }{\cal B}i_{\bar \eta }d{\cal A}\nonumber \\
                    &\approx & 0, \label{eq:2.29}
\end{eqnarray}
the two terms being constraints (of types ${\cal G}_0$ and ${\cal G}_1$).

We have now established that $l(\eta )$ is an edge observable creating edge
diffeos.

The Fourier analysis of edge observables is important for quantization and
useful for establishing that $\l(\eta )$ has the generalized Sugawara form. For
performing this analysis, we must first fix a volume form $\mu $ on $\partial
\Sigma $. For $\partial \Sigma =S^2$, it can for instance be constant$\times
d(\cos \theta )d\varphi $, with $\theta $ and $\varphi $ being polar and
azimuthal angles. For $\partial \Sigma =T^2$, it can be ${\rm constant}\times
d\theta ^1d\theta ^2$, $\theta ^i$ (mod $2\pi $) being angular coordinates.
The choice of $\mu $ is not of course  unique. It defines a Hilbert space
${\cal H}=L^2(\mu ,\partial \Sigma )$ of square integrable functions with
respect to $\mu $. Thus, if $\psi ,\, \chi \in {\cal H}$, then
\be
\left( \psi ,\chi \right) \equiv \int _\Sigma \mu \psi ^{*}\chi <\infty .
\label{eq:2.30}
\ee

Let $e_n$ be an orthonormal basis of smooth functions for ${\cal H}$ with $e_0$
the constant function:
\be
\left( e_n,e_m\right) =\delta _{nm},\,\,e_0={\rm constant}\,.
\label{eq:2.31}
\ee
For $S^2$, for
\be
\mu =\frac A{4\pi ^2}d(\cos \theta )d\varphi \label{eq:2.32}
\ee
[with $\theta $ and $\varphi $ being the usual polar and azimuthal angles],
we can assume the correspondences
$$n\rightarrow Jm,$$
\be
e_n\rightarrow \left( \frac {4\pi }A\right)
^{1/2}Y_{Jm},\,\,\,\,Y_{Jm}={\rm Spherical\,\, harmonics},\label{eq:2.33}
\ee
$e_0$ becoming $\left( \frac{4\pi }A\right) ^{1/2}$.
For $T^2$, for
\be
\mu =\frac A{4\pi ^2}d\theta ^1d\theta ^2 ,\label{eq:2.34}
\ee
we can assume the correspondences
$$
n\rightarrow \vec N\equiv (N_1,N_2),
$$
$$e_n(\theta ^1,\theta ^2)\rightarrow e_{\vec N}= \frac 1{\sqrt A}
e^{i\vec N\cdot\vec \theta },
$$
\be
\vec N\cdot \vec \theta \equiv N_1\theta ^1+N_2\theta ^2,\,\,\,N_i\in Z,
\label{eq:2.35}
\ee
$e_0$ becoming $\frac {1}{\sqrt A}$.
In both these cases, $A$ is the area of $\partial \Sigma $:
\be
 \int _{\partial \Sigma }\mu =A .
\label{eq:2.36}\ee

With this setup, the observables can be Fourier analyzed. Consider first
$q(d\xi ^{(0)})$. All $d\xi ^{(0)}$ with equal boundary value $\xi
^{(0)}|_{\partial \Sigma }$ give equivalent observables. Let $\langle q(d\xi
^{(0)})\rangle $ denote this equivalence class. We will hereafter call it as
a single ($q$ type) observable.
It depends only on $\xi ^{(0)}|_{\partial \Sigma }$.
A basis of such $q$ type observables is thus obtained by
choosing $\xi ^{(0)}|_{\partial \Sigma }$ to be $e_n$:
\be
q_n=\langle q(d\xi _n^{(0)})\rangle ,\,\,\,\,\,\,\,\, \xi
_n^{(0)}|_{\partial \Sigma }=e_n,\,\, n\neq 0.
\label{eq:2.35-a}\ee
$<q(d\xi ^{(0)}_0>$ here is the null observable. This is because
$\xi _0^{(0)}|_{\partial
\Sigma }$, being a constant, can be extended as a constant function to all of
$\Sigma $, and in that case $d\xi _0^{(0)}=0$. It is for this reason that we
have excluded $n=0$ from (\ref{eq:2.35-a}).

In addition to $q_n$, there is also one more $q$ type observable $Q$. It is
the equivalence class of observables weakly equal to
\be
\frac 1{\sqrt A}q(d\theta ^1) .\label{eq:2.35-iii}
\ee
Note that $Q\neq \langle q(d\xi ^{(0)}_0)\rangle $, the equivalence class of
observables weakly equal to $q(d\xi ^{(0)}_0)$.

As for $p(d\xi ^{(1)})$, let $\langle p(d\xi ^{(1)})\rangle $ denote the
equivalence of weakly equal observables, all with the same $d\xi
^{(1)}|_{\partial \Sigma }$. We will hereafter call it a single ($p$ type)
observable. We can choose $e^{*}_n\mu $ ($n\neq 0$) for
$d\xi ^{(1)}|_{\partial \Sigma }$ to obtain a class of $p$ type observables,
$e^{*}_n$ being the complex conjugate of $e_n$.
This is because $e^{*}_n\mu $ integrates
to zero on $\partial \Sigma $ if $n\neq 0$, and for $S^2$ or $T^2$, this means
that it is exact, $e_n^{*}\mu =d\chi _n^{(1)}$. But $\chi_n^{(1)}$ can always
be extended (in fact in many ways) from $\partial \Sigma $ to
a form $\xi ^{(1)}_n$ on $\Sigma $. Hence the choice $\xi _n^{(1)}$ for $\xi
^{(1)}$ gives us $e^{*}_n\mu $ for $d\xi _n^{(1)}|_{\partial \Sigma }$. We can
thus take
\be
p_n=\langle p(d\xi _n^{(1)})\rangle ,\,\,\,\,\,\,\,\,\,\,\,\,\,\,\,n\neq 0
\label{eq:2.36-a}\ee
as a class of $p$ type observables.

The observables (\ref{eq:2.36-a}) form a basis for $p$ type observables for a
ball. For a solid torus, to obtain a basis for $p$ type observables, we must
also consider, for example,
$$
p(d\psi ^{(1)})\,\,,
$$
\be
\psi ^{(1)}={\rm Any\,\,one\,\,form\,\,fulfilling\,\,} \psi ^{(1)}|_{\partial
\Sigma }=-\frac {\sqrt A}{4\pi ^2}d\theta ^2. \label{eq:2.39-iii}
\ee

Let $P$ denote the equivalence class of observables of the type
(\ref{eq:2.39-iii}), all with the same $\psi ^{(1)}|_{\partial \Sigma }$.
Then $P$ and $p_n$ form a basis for $p$ type observables for ${\bf T}_3$.
The observable $P$ is missing from the set $\lbrace p_n \rbrace $ in
(\ref{eq:2.36-a}) because $d\psi ^{(1)}|_{\partial \Sigma }$=0. [Note that
$P\neq \langle p(d\xi ^{(1)}_0)\rangle $, the equivalence class of observables
weakly equal to $p(d\xi ^{(1)}_0)$.]

The PB's of these observables are determined by (\ref{eq:2.20}) and
(\ref{eq:2.22-iii}). The nonzero PB's which involve them are given by
\begin{eqnarray}
\left\{ q(d\xi _m^{(0)}),p(d\xi _m^{(1)})\right\}
                                     & = & \int _{\partial \Sigma }\xi
_m^{(0)}d\xi _n^{(1)}
\nonumber \\
                                     & = & \delta _{nm},\nonumber
\end{eqnarray}
\begin{eqnarray}
\left\{ \frac 1{2\pi }q(d\theta ^1),p(d\psi ^{(1)})\right\}
& = & \frac 1{4\pi ^2}\int _{\partial \Sigma }d\theta ^1d\theta ^2
\nonumber \\
& = & 1\nonumber
\end{eqnarray}
and read
$$
\left\{ q_m,p_n\right\}=\delta _{mn}\,,
$$
\be
\left\{ Q,P\right\}=1\,.\label{eq:2.37}
\ee

There is an interpretation of the observables $Q$ and $P$ which will be briefly
alluded to here, a fuller discussion being reserved for ref.\cite{we}. The
observable $Q$ is associated with the operator which creates magnetic flux
loops which loops are homologous to the cycle on $\partial \Sigma $ obtained
by varying $\theta ^2$ from $0$ to $2\pi $ with fixed $\theta ^1$. The value
of the observable $P$ is a measure of the flux on these loops.

Now for the Fourier components of the equivalence class $\langle l(\eta
)\rangle $ of observables, all with the same $\eta |_{\partial \Sigma }$, we
adopt the choices (\ref{eq:2.32})-(\ref{eq:2.36}). We may then set
\be
\eta =\eta _{Jm,\alpha }\,\,,\label{eq:2.38-a}
\ee
\be
\eta _{Jm,\alpha }|_{\partial \Sigma }=Y_{Jm}L_\alpha \label{eq:2.38}
\ee
for $S^2$, $L_\alpha $ being the angular momentum generators:
\be
{\cal L}_{L_\alpha }\mu =0,\,\,\,\,\,\,\,\,\,\,\left[ L_\alpha ,L_\beta
\right] =\epsilon _{\alpha \beta \gamma }L_\gamma .\label{eq:2.39}
\ee
(\ref{eq:2.38}) is permissible as its right hand side can be extended to all of
$\Sigma $. The corresponding $\langle l(\eta _{Jm,\alpha })\rangle $'s are
\be
l_{Jm,\alpha }=\langle l(\eta _{Jm,\alpha })\rangle .\label{eq:2.40}
\ee

For $\partial \Sigma = T^2$, we can analogously set
     $$\eta =\eta _{\vec N,j}$$
               $$\eta _{\vec N,j}\left| _{\partial \Sigma }\right. =e^{i\vec
N\cdot \vec \theta
}\frac \partial {\partial \theta ^j},$$
$${\cal L}_{\frac \partial {\partial \theta ^j}}\mu =0\,\,,$$
\be
l_{\vec N,j}=\langle l(\eta _{\vec N,j})\rangle .
\label{eq:2.41}
\ee

We next give the PB's involving $l_{\vec N,j}$. Those involving
(\ref{eq:2.40}) are also straightforward to derive, but involve Clebsch-Gordan
coefficients and are omitted for simplicity. We have,
$$
\left\{ l_{\vec N,j},q_{\vec M}\right\}  = iM_jq_{\vec M+\vec N}\,\,,
$$
$$
\left\{ l_{\vec N,j},Q\right\}=\delta _{j1}q_{\vec N}\,,
$$
$$
\left\{ l_{\vec N,j},p_{\vec M}\right\}  =-i(M_j - N_j)p_{\vec M - \vec N}
-\delta _{j1}\delta _{\vec N,\vec M}P\,,
$$
$$
\left\{ l_{\vec N,j},P\right\}=0\,,
$$
\be
\left\{ l_{\vec N,j},l_{\vec M,k}\right\}=iM_jl_{\vec N +\vec M,k} -
iN_kl_{\vec N + \vec M,j}\,\,.\label{eq:2.42}
\ee
where $\delta _{\vec N,\vec M}=\delta _{N_1,M_1}\delta _{N_2,M_2}$. The
derivations of only the second, the last term in the third, and the fourth
equations merit comment. The second equation follows easily from the first
equation of (\ref{eq:2.22}) [with $w^{(1)}=\frac 1{\sqrt A}d\theta ^1$] and
(\ref{eq:cartan}).
As for the last term in the third
equation, we
have, $\left\{ l(\eta _{\vec N,j}),p(d\xi ^{(1)}_{\vec N})\right\}=
p\left( d\left\{ i_{\eta _{\vec N,j}}\xi ^{(1)}_{\vec N}\right\}\right)$
after using
(\ref{eq:cartan}).
Now $i_{\eta _{\vec N,j}}
\left. \xi ^{(1)}_{\vec N}\right|_{\partial \Sigma } =\frac {\sqrt A}{4\pi ^2}
\left[ \delta _{j1}
d\theta ^2-\delta _{j1}d\theta ^1\right] $. The second term here can be
extended
to all of $\Sigma $ as a closed form and thus contributes nothing. The first
term
gives the term in question. As for the fourth equation, one has,
$\left\{ l(\eta _{\vec N,j}),p(d\psi ^{(1)})\right\}=p\left(
d{\cal L}_{\eta _{\vec N,j}}\psi ^{(1)}\right)$. Now $\left( {\cal L}
_{\eta _{\vec N,j}}\psi ^{(1)}\right)\left| \right. _{\partial \Sigma } =
{\cal L}_{\eta _{\vec N,j}}\left| \right. _{\partial \Sigma }\left( -\frac
{\sqrt A}{4\pi ^2}d\theta ^2\right)$ which on using
(\ref{eq:cartan})
becomes
$d\left( -
\frac {\sqrt A}{4\pi ^2}e^{i\vec N.\vec \theta }\delta _{j2}\right)$. This
can be extended as an exact form $\chi ^{(1)}$ to all of $\Sigma $ so that
$p(d\chi ^{(1)})=0$ (in a trivial way). The fourth equation above follows
readily.

We next note that the observables
\be
\hat l_{\vec N,j}=-i\sum _{\vec M}M_jq_{\vec M +\vec N}p_{\vec M}-\delta _{j1}
q_{\vec N}P \label{eq:2.43}
\ee
have the same PB's as $\l_{\vec N,j}$. It must therefore the case that
\be
l_{\vec N,j}= \hat l_{\vec N,j}
\label{eq:2.44}
\ee
which is the generalized classical Sugawara formula.

We can show (\ref{eq:2.44}) explicitly in the following way. Choose a
Euclidean metric on (the interior of) ${\bf T}_3$. Relative to this metric,
let $r$ be the
radial distance from the central thread of the solid torus ${\bf T}_3$
with $\partial
\Sigma $ having $r=1$. Then $(r,\theta ^1,\theta ^2)$ are coordinates for
${\bf T}_3$. Let $\Lambda $ be a function of $r$ alone with
\be
\Lambda (r)=0,\,{\rm for}\,\,r<\epsilon <1\,,\,\,\,\,\Lambda (1)=1,\,\,\,\,
\,\,d\Lambda (1)=0\,, \label{eq:2.45}
\ee
$\epsilon $ being a small positive number.
We may then set
\be
\eta _{\vec N,j}(r,\theta ^1,\theta ^2)=\Lambda (r)e^{i\vec N\cdot \vec
\theta }\frac \partial {\partial \theta ^j}\,\, . \label{eq:2.46}
\ee
For this choice,
\begin{eqnarray}
l_{\vec N,j}&=& \int _{\partial \Sigma }e^{i\vec N\cdot \vec \theta }
{\cal A}_j(1,\vec \theta ) {\cal B}(1,\vec \theta )
-\int _\Sigma \left(\Lambda (r)e^
{i\vec N\cdot \vec \theta }{\cal A}_j(r,\vec \theta )\right)d{\cal B} +
\nonumber \\
&  & +\int _\Sigma \Lambda (r)e^{i\vec N\cdot \vec \theta }{\cal B}(\vec y)
{\cal F}_{j\alpha }(\vec y)dy^\alpha \label{eq:2.47}
\end{eqnarray}
where ${\cal F}_{j\alpha }= \partial _j{\cal A}_\alpha - \partial _\alpha
{\cal A}_j$.[There is a slight (and inconsequential) abuse of notation in
(\ref{eq:2.47}) which is really $l(\eta _{\vec N,j})$ and not $l_{\vec N,j}$.
There are similar unimportant inaccuracies in what follows.]

In (\ref{eq:2.47}) and in the equations up to (\ref{eq:2.53}), the components
of forms and vector fields are with reference to the coordinate system
$(r,\theta ^1,\theta ^2)\equiv (y^1,y^2,y^3)\equiv \vec y$. [Elsewhere, we
have instead
used Cartesian coordinates.] The Levi-Civita symbols $\epsilon ^{\alpha \beta
\gamma } [1\leq \alpha ,\beta ,\gamma \leq 3]$ and $\epsilon _{jk}[2\leq j,k
\leq 3]$ for these equations are so defined that
$\epsilon ^{123}=1=\epsilon _{23}$. Also ${\cal B}(1,\vec \theta )$ denotes
${\cal B}_{jk}(1,\vec \theta )d\theta ^jd\theta ^k$.

Let
$$
\Lambda e_{\vec M},
$$
\be
\frac A{4\pi ^2}\Lambda \frac {e^{*}_{\vec M}}{-iM_j}\epsilon _{jk}d\theta ^k
\,\,\,\, {\rm (no\,\,}j{\rm \,\,summation)}\label{eq:2.48}
\ee
be the extensions of $\xi _{\vec M}^{(0)}|_{\partial \Sigma }=e_{\vec M}$
and $\xi _{\vec M}^{(1)}|_{\partial \Sigma }$ to all of
$\Sigma $. We then have
$$
q_{\vec M}=\int _{\partial \Sigma }e_{\vec M}{\cal B} - \int _\Sigma \Lambda
e_{\vec M}d{\cal B},
$$
\be
-iM_jp_{\vec M}=\int _{\partial \Sigma }{\cal A}_je^{*}_{\vec M}\mu -
\frac {A}{4\pi ^2} \int _\Sigma \Lambda e^{*}_{\vec M}\epsilon _{jk}
d\theta ^kd{\cal A}\,,\,\,M_j\neq 0\,.\label{eq:2.49}
\ee

We next examine
$$
-i\sum _{{\vec M\,\,{\rm with}}\atop M_j \neq 0} M_jq_{\vec M+\vec N}
p_{\vec M}
$$
$$
-\frac A{4\pi ^2}\sum _{{\vec M\,\,{\rm with}}\atop M_j = 0}
\int _\Sigma d\left[ \Lambda (r')
e_{\vec M+\vec N}(\vec \theta ')\right]{\cal B}(r',\vec \theta ')\int _
\Sigma d\left[ \Lambda (r)e^*_{\vec M}(\vec \theta )\epsilon _{jk}d\theta ^k
\right] {\cal A}(r,\vec \theta )
$$
\be
=-i\sum _{{\rm All}\,\,\vec M}\left[ \int _{\partial \Sigma }
e_{\vec M+\vec N}{\cal B}-\int _\Sigma \Lambda e_{\vec M+\vec N}d{\cal B}
\right] \left[ \int_{\partial \Sigma }{\cal A}_je^*_{\vec M}\mu
-\frac A{4\pi ^2}\int _\Sigma \Lambda e^*_{\vec M}\epsilon _{jk}d\theta ^kd
{\cal A} \right] \,.\label{eq:2.56-v}
\ee

Consider the second term on the left hand side of (\ref{eq:2.56-v}). We will
now show that it is the same as the last term on the right hand side of
(\ref{eq:2.43}). First consider $j=2$. For this $j$, the second integral in
this term is
\be
-\frac 1{\sqrt A}\int _\Sigma d\left[ \Lambda (r)e^{-iM_1\theta ^1}d\theta ^1
\right]{\cal A}(r,\vec \theta )\,. \label{eq:2.57-v}
\ee
$e^{-iM_1\theta ^1}d\theta ^1$ being a globaly defined closed one form on
${\bf T}_3$, this is equal to
\be
-\frac 1{\sqrt A}\int _\Sigma d\left[ (\Lambda (r)-1)e^{-iM_1\theta ^1}
d\theta ^1\right]{\cal A}(r,\vec \theta )\,. \label{eq:2.58-v}
\ee
As
\be
(\Lambda (r)-1)e^{-iM_1\theta ^1}d\theta ^1\left| \right. _{\partial \Sigma }
=0\,, \label{eq:2.59-v}
\ee
this expression, and hence the term in question, are weakly zero. This is
what we want, the last term in (\ref{eq:2.43}) being zero for $j=2$.

For $j=1$, the analogue of (\ref{eq:2.57-v}) is
\be
\frac 1{\sqrt A}\int _\Sigma d\left[ \Lambda (r)e^{-iM_2\theta ^2}d\theta ^2
\right]{\cal A}(r,\vec \theta ) \label{eq:2.60-v}
\ee
which for $M_2\neq 0$ is
\be
\frac 1{\sqrt A}\int _\Sigma d\left[ \Lambda (r)e^{-iM_2\theta ^2}d\theta ^2
-d\left( \frac {\Lambda (r)e^{-iM_2\theta _2}}{-iM_2}\right)
\right]{\cal A}(r,\vec \theta )\,. \label{eq:2.61-v}
\ee
Since
\be
\left. \left[ \Lambda (r)e^{-iM_2\theta ^2}d\theta ^2-d\left( \frac
{\Lambda (r) e^{-iM_2\theta _2}}{-iM_2}\right)\right] \right| _{\partial
\Sigma }=0 \label{eq:2.62-v}
\ee
in view of (\ref{eq:2.45}), this term is weakly zero. When both $M_1$ and
$M_2$ are
zero, the second term on the left hand side of (\ref{eq:2.56-v}) gives just
$-q_{\vec N}P$. Thus the left hand side of (\ref{eq:2.56-v}) is the same as
the right hand side of (\ref{eq:2.43}), that is $\hat l_{\vec N,j}$.

The right hand side of (\ref{eq:2.56-v}) gives
\begin{eqnarray}
\hat l_{\vec N,j}&=&\int _{\partial \Sigma }e^{i\vec N\cdot \vec \theta}
{\cal A}_j(1,\vec \theta ){\cal B}(1,\vec \theta )
+\int _\Sigma \Lambda (r)e^{i\vec N\cdot \vec \theta }{\cal B}
(1,\vec \theta){\cal F}_{j\alpha }(\vec y)dy^\alpha  \nonumber \\
& & -\int _{\Sigma }\Lambda (r)e^{i\vec N\cdot \vec \theta }{\cal A}_j(1,
\vec \theta )d{\cal B}(r,\vec \theta )\nonumber \\
& & +\int _\Sigma \Lambda (r,\vec \theta )e^{i\vec N\cdot \vec \theta }\left[
\int^\infty _0 dr'\Lambda (r')\partial _\lambda{\cal B}_{\rho \sigma }
(r',\vec \theta )\epsilon ^{\lambda \rho \sigma }\right]\epsilon _{jk}
d\theta ^kd{\cal A}(r,\vec \theta) \label{eq:2.50}
\end{eqnarray}
where the completeness relation
\be
\sum _{\vec N}e^{i\vec N\cdot (\vec \theta -\vec \theta ^{\prime
})}=4\pi ^2\delta (\theta _1-\theta _1^{\prime })\delta (\theta _2
-\theta _2^{\prime }) \label{eq:2.51}
\ee
has been used.

The test function for the Gauss law generator $d{\cal A}$ in the last term
involves the Gauss law generator $d{\cal B}$. We therefore interpret it as
weakly zero and get
\begin{eqnarray}
\hat l_{\vec N,j} & = & \int _{\partial \Sigma }e^{i\vec N\cdot \vec
\theta}{\cal A}_j(1,\vec \theta ){\cal B}(1,\vec \theta ) + \int _\Sigma
\Lambda (r)e^{i\vec N\cdot \vec \theta }{\cal B}(1,\vec \theta )
{\cal F}_{j\alpha }(\vec y)dy^\alpha \nonumber \\
& & - \int _\Sigma \Lambda (r)e^{i\vec N\cdot \vec \theta }{\cal A}_j
(1,\vec \theta )d{\cal B}(r,\vec \theta )\,. \label{eq:2.52}
\end{eqnarray}

Now we can write ${\cal BF}_{j\alpha }dy^\alpha $ as $-2{\cal B}_{j\alpha }
dy^\alpha
d{\cal A}$. Also,in view of (\ref{eq:2.45}), and the fact that $j$ refers to
components tangent to $\partial \Sigma $ when $r=1$, it is clear that
$$
\left[\Lambda (r)e^{i\vec N\cdot \vec \theta }{\cal A}_j(r,\vec \theta )-
\Lambda (r)e^{i\vec N\cdot \vec \theta }{\cal A}_j(1,\vec \theta )\right]
\mid _{\partial \Sigma }\,\,= 0\,,
$$
\be
\left[\Lambda (r)e^{i\vec N\cdot \vec \theta }{\cal B}_{j\alpha }(\vec y)
dy^\alpha - \Lambda (r)e^{i\vec N\cdot \vec \theta }{\cal B}_{j\alpha }
(1,\vec \theta)d\theta ^ \alpha \right]\mid _{\partial \Sigma }\,\,=0 .
\label{eq:2.53}
\ee
We thus find (\ref{eq:2.44}).

For $\partial \Sigma =S^2$, we can find the analogues
$\hat l_{Jm,\alpha }$ of (\ref{eq:2.43}) which
have the same PB's as $l_{Jm,\alpha }$. This leads to the generalized
classical Sugawara formula
\be
\hat l_{Jm,\alpha }=l_{Jm,\alpha }.
\label{eq:2.55}
\ee
[We do not reproduce the formula for $\hat l_{Jm.\alpha }$ as it is
complicated and not very illuminating.]
It should be possible to verify (\ref{eq:2.55}) explicitly as for the torus,
although we have not done so.

\newpage
\sxn{THE TOPOLOGICAL $BF$ THEORY}\label{sec-top.bf}

In this Section, we show that when the system of Section 2 is in its
ground state, it is described by the topological $BF$ field theory. The
latter also predicts edge states with properties similar to those of the last
Section.

For the Hamiltonian (\ref{eq:2.4}), the energy density in $\Sigma ^0$ is zero
if
$$
\pi _i-\epsilon _{ijk}B_{jk}=0,\,\,\,\,\,P_{ij}=0,
$$
\be
F_{ij}=0,\,\,\,\,\,\,\,H_{ijk}=0.\label{eq:3.1}
\ee

The symplectic one form
$$\theta =\int \left( \pi _i\delta A_i+\frac 12P_{ij}\delta B_{ij}\right)
d^3x$$
appropriate to Section 2 restricted (pulled back) to the
surface (\ref{eq:3.1}) becomes
\be
\theta ^{*}=\int \epsilon _{ijk}B_{jk}\delta A_id^3x ,\label{eq:3.2}
\ee
while the constraints ${\cal G}_i$ pulled back to (\ref{eq:3.1}) become
$${\cal G}_0^{*}=\int d^3x\lambda ^{(0)}dB\approx 0\,\,,$$
\be
{\cal G}_1^{*}=2\int d^3x\lambda ^{(1)}dA\approx 0\,\,.
\label{eq:3.3}\ee

We can also pull back the observables $q(w^{(1)}),\,p(w^{(2)})$ and
$l(\eta )$. They become
$$q^{*}(w^{(1)})=\int w^{(1)}B,$$
$$p^{*}(w^{(2)})=-\int w^{(2)}A,$$
\be
l^{*}(\eta )=\int ({\cal L}_\eta A)B.\label{eq:3.4}
\ee

The PB's of $A$ and $B$  follows from (\ref{eq:3.2}):
$$\left\{ A_i(x),A_j(y)\right\} =\left\{ B_{ij}(x),B_{kl}(y)\right\} =0,$$
\be
\left\{ B_{ij}(x),A_k(y)\right\} =-\frac 12\epsilon _{ijk}\delta ^3(x-y).
\label{eq:3.5}\ee
All fields here are evaluated at equal times.

It is to be observed that (\ref{eq:3.3})-(\ref{eq:3.5}) can be obtained from
Section 2 by substituting $A$, $B$ for ${\cal A}$, ${\cal B}$.
The entire previous description of constraints and edge excitations can
therefore
be transferred intact to the surface (\ref{eq:3.1}).

It is also to be observed that (\ref{eq:3.2})-(\ref{eq:3.5}) are consequences
of the topological $BF$ action
\be
S^{*}=\int dtL^{*}
\label{eq:3.6}\ee
of the Lagrangian (\ref{eq:1.7}). Thus when a system described by London
equations is in its ground state, its edge excitations are described by a
topological field theory.

Henceforth, we will work with the Lagrangian (\ref{eq:1.6}), but it will be
obvious that we could equally well have worked with the topological field
theory.

\newpage
\sxn{QUANTIZATION AND THE GROUP $SDIFF(\partial \Sigma
)$}\label{sec-quantization}

We here consider $\Sigma =B_3$ or ${\bf T}_3$. The quantum operators for
$q.,\,p.,\,l.\,$ will be denoted by the corresponding capital letters. The
quantum operators for $Q$ and $P$ will be denoted by ${\cal Q}$ and
${\cal P}$.We will also adopt the choice (\ref{eq:2.30})-(\ref{eq:2.36}) for
describing the Fourier components.

Now since
\be
Y^{*}_{Jm}=(-1)^mY_{J,-m},\,\,\,\,\left( e^{i\vec N\cdot \vec \theta }\right)
^{*}=e^{-i\vec N\cdot \vec \theta },
\label{eq:4.1}
\ee
we have
$$
Q_{Jm}^{\dag }=(-1)^mQ_{J,-m},\,\,\,\,\,\,\,P_{Jm}^{\dag } =(-1)^mP_{J,-m},
$$
\be
Q^{\dag }_{\vec N}=Q_{-\vec N}\,,\,\,\,\,\,\,\,\,\,\,\,\,\,\,P^{\dag }_{\vec
N}=P_{-\vec N}\,\, .\label{eq:4.2}
\ee

Let $\omega :n\rightarrow \omega (n)\,(>0)$ be a frequency function invariant
under the substitution
$$n=Jm\rightarrow n^{*}\equiv J,-m$$
or
\be
n=\vec N\rightarrow n^{*}\equiv -\vec N. \label{eq:4.3}
\ee
The dispersion relation is otherwise left arbitrary for the moment.

For the moment, let us set aside the modes ${\cal Q}$ and ${\cal P}$ which
exist for ${\cal T}_3$.

We now form the annihilation and creation operators
$$a_n=\frac 1{\sqrt 2}\left[ \omega (n)Q_n+iP^{\dag }_n\right] ,$$
\be
a^{\dag }_n=\frac 1{\sqrt 2}\left[ \omega (n)Q^{\dag }_n-iP_n\right] .
\label{eq:4.3-b}\ee
Their only nonzero commutator is
\be
\left[ a_n,a_m^{\dag }\right] =\omega (n)\delta _{nm}\,.
\label{eq:4.4}\ee

The Fock space quantization of (\ref{eq:4.4}) is standard. Let $|0\rangle $
denote the Fock space vacuum:
\be
a_n|0\rangle =0\,.
\label{eq:4.5}\ee

The quantum version $L_n$ of diffeo generators are obtained from their
classical expressions after normal ordering. We now argue the following: a) Not
all $L_n$ can be implemented on the preceding Fock space regardless of the
dispersion relation . b) Let $SDiff_0(\partial \Sigma )$ denote the group of
diffeos leaving $\mu $ invariant with classical generators ${s_n}$. Their
quantum versions $S_n$ can be implemented on the preceding Fock space if
$\omega (n)$ is independent of $n$.

In this way, by demanding the implementability of $S_n$, we gain some control
over the dispersion relation just as in the Chern-Simons case. As remarked in
the Introduction, this requirement will also suggest interesting field
theories for describing the edge excitations.

As for a), consider for example the squared norm
\be
{\cal N}^2=\left\langle 0\left| L_{\vec N,j}^{\dag }L_{\vec N,j}\right|
0\right\rangle
\label{eq:4.6}
\ee
of the state $L_{\vec N,j}\left| 0\right. \left\rangle {}\right. $ for
$\partial \Sigma =T^2$. Here
\be
L_{\vec N,j}=-\frac 12:\sum _{\vec M}\frac {M_j}{\omega (\vec M+\vec N)}
(a_{\vec M+\vec N}+ a^{\dag }_{-\vec M-\vec N})
(a_{-\vec M}-a^{\dag }_{\vec M}):\label{eq:4.7}
\ee
where we have ignored a term containing ${\cal P}$.

The computation of (\ref{eq:4.6}) requires regularisation. We interpret it as
$$\lim _{k\rightarrow \infty }{\cal N}_k^2\,\, ,$$
\be
{\cal N}_k^2=\left\langle 0\left| L_{\vec N,j}^{(k){\dag }}L_{\vec
N,j}^{(k)}\right| 0\right\rangle
\label{eq:4.8}
\ee
where
\be
L_{\vec N,j}^{(k)}=-\frac 12\sum _{{|M_i|<k}\atop for\,\,i=1,2}\frac {M_j}
{\omega (\vec M+\vec N)}
:(a_{\vec M+\vec N}+a^{\dag }_{-\vec M-\vec N})(a_{-\vec M}-
a^{\dag }_{\vec M}): \label{eq:4.9}
\ee
Substitution of (\ref{eq:4.9}) in (\ref{eq:4.8}) shows that
$$
{\cal N}_k^2=\frac 14\sum _{{|M_i|<k}\atop for\,\,i=1,2}M_j^2\frac {\omega
(\vec M)}{\omega (\vec M+\vec N)}-\frac 14\sum _{{|M_i|\,{\rm and}\,
|\bar M_i|<k}\atop for\,\,i=1,2}
\left( M_j^2+N_jM_j\right)\prod _{l=1,2}\delta _{\bar M_l,-M_l-N_l}
$$
If both $N_i$ for example are positive, this becomes
\be
{\cal N}_k^2=\frac 14\sum _{{|M_i|<k}\atop for\,\,i=1,2}M_j^2\frac {\omega
(\vec M)}{\omega (\vec M+\vec N)}-\frac 14\sum _{M_1=-k}^{k-N_1}
\sum _{M_2=-k}^{k-N_2}\left( M^2_j+N_jM_j\right).\label{eq:4.10}
\ee
We can find no function $\omega $ for which this expression is finite as
$k\rightarrow \infty $ and therefore (tentatively) conclude that there is
no choice of $\omega $ for which $L_{\vec N,j}$ is well defined on our Fock
space.

Now there is a local scalar field Lagrangian invariant under $SDiff(\partial
\Sigma )$ and with a prescribed dispersion relation, namely (\ref{eq:1.8}).
It turns out that the Lie algebra of the group $SDiff_0(\partial \Sigma )$ is
implementable by operators in the quantum theory of this Lagrangian. We
shall now study this Lagrangian, argue that it can describe the edge states
and finally show the implementability of the algebra of the group
$SDiff_0(\partial \Sigma )$ in the quantum theory of (\ref{eq:1.8}).

The field $\varphi $ described by (\ref{eq:1.8}) is characterized by the
frequency $\omega (n)=\omega _0$ independent of $n$. It becomes a quantum
field $\Phi $ if we set
$$\Phi =\frac 1{\sqrt {2\omega _0}}\sum _{n}\left( a_ne_n+
a_n^{\dag }e_n^{*}\right) $$
\be
\dot \Phi =-i\sqrt {\frac {\omega _0}2}\sum _{n}\left( a_ne_n-a_n^{\dag }
e_n^{*}\right)\label{eq:4.10-a}
\ee
where
\be
\left[ a_n,a^{\dag }_m\right] =\delta _{nm}\label{eq:4.11}
\ee
and define the vacuum $\left| 0\right. \left\rangle {}\right. $ by
\be
a_n\left| 0\right. \left\rangle {}\right. =0\label{eq:4.12}
\ee
Its Hamiltonian has the expression $\sum \omega _0a_n^{\dag }a_n$.

In comparison with the previous $a_n$, the field $\Phi $ also has the modes
$a_0,\,a_0^{\dag }$. They can be eliminated by considering
$L_\alpha \Phi $(for $S^2$) or $\frac \partial {\partial \theta ^j}\Phi $ (for
$T^2$). Thus the modes we find are those of these gradient fields. In the same
way, the Chern-Simons theory on a disc $D$ describes a suitable derivative of a
scalar field on the boundary $\partial \Sigma $ \cite{bbgs1}. As in that
theory, these zero modes become relevant when $\partial \Sigma $ has more than
one connected component or there are sources, when $a_0$ and $a_0^{\dag }$
acquire an interpretation in terms of electric charge excitations, $\frac 1
{\sqrt 2 i}(a_0-a_0^{\dag })$ being the charge operator. We will show these
results in a second paper \cite{we}. But there is a problem with the
Lagrangian (\ref{eq:1.8}) when the zero modes are important. The zero mode
part $\omega _0a_0^{\dag }a_0$ of its Hamiltonian is not diagonal when charge
is diagonal. For this reason, (\ref{eq:1.8}) must be used with caution when
zero modes are significant. We emphasize that (\ref{eq:1.8}) has been used
here only to motivate our conclusion that $SDiff_0(\partial \Sigma )$ is
implementable for the edge states of (\ref{eq:1.6}) and (\ref{eq:1.7}). None
of our substantive results depend crucially on the use of (\ref{eq:1.8}).

We have yet to attend to the modes ${\cal Q}$ and ${\cal P}$. As we will argue
in \cite{we}, ${\cal P}$ measures the magnetic flux tangent to
$\partial \Sigma $ (such as that in vortices winding around $\partial \Sigma
$) and ${\cal Q}$ is associated with the creation of such flux. We will
regard ${\cal P}$ as a superselected operator with eigenvalue $\Theta $ for
state $\left| 0\right. \rangle $ and ${\cal Q}$ as associated with vertex
operators for the creation of such flux. This assumption is similar to the
one
adopted in Chern-Simons dynamics or conformal field theories in the
treatment of charge or `momentum' \cite{bbgs1,bbgs2,goddard}. Thus $\left| 0
\right. \rangle $ is a unit norm state with ${\cal P}\left| 0\right.
\rangle =\Theta \left| 0\right. \rangle $. Note that the inclusion of the term
$-\delta _{j1}\frac 1{\sqrt 2\omega (\vec N)}[a^{\dag }_{\vec N}-a_{-\vec N}]
{\cal P}$, ignored in (\ref{eq:4.7}), does not affect our conclusion regarding
the divergence of ${\cal N}^2$.

The following vector fields $\eta |_{\partial \Sigma }$ preserve $\mu $:
$$
S^2:\,\,\,\,\,\,\,\,\,\eta \mid _{\partial \Sigma }=v_{lm},
$$
\be
v_{lm}=\frac {\partial Y_{lm}}{\partial (cos\theta )}\frac {\partial }
{\partial \phi}-\frac {\partial Y_{lm}}{\partial \phi}\frac {\partial}
{\partial (cos\theta )}\label{eq:4.13}
\ee
\begin{eqnarray}
T^2:\,\,\,\,\,\,\,\,\eta \mid _{\partial \Sigma }&=&v_{\vec N}\nonumber \\
{\rm or}\,\,\,\,\,\,\eta \mid _{\partial \Sigma }&=&t_i\,\, ,\nonumber
\end{eqnarray}
\begin{eqnarray}
v_{\vec N}&=& \frac {\partial e^{*}_{\vec N}}{\partial \theta ^1}
\frac {\partial }{\partial \theta ^2}-\frac {\partial e^{*}_{\vec N}}
{\partial \theta ^2}\frac {\partial }{\partial \theta ^1} \nonumber \\
& = & -ie_{\vec N}^{*}\left( N_1\frac {\partial }{\partial \theta ^2}-
N_2\frac {\partial }{\partial \theta ^1}\right), \nonumber
\end{eqnarray}
\be
t_i=\frac {\partial }{\partial \theta ^i}\,\, .\label{eq:4.14}
\ee

For the field $\varphi $, the classical generators of transformations due to
these vector fields are
\be
T^2:\,\,\,\,s^\varphi _\eta =\int \mu \varphi {\cal L}_{\eta }\dot \varphi
\label{eq:4.16}
\ee
and their quantum versions for the torus are
\be
{\cal S}^\Phi _{\vec N}=\sum _{\vec M}\frac i2(\vec N\times \vec M):\left(
a_{\vec M+\vec N}+a^{\dag }_{-\vec M-\vec N}\right) \left( a_{-\vec M}-
a^{\dag }_{\vec M}\right) :\,\,{\rm for}\,\, \eta |_{\partial \Sigma }=
v_{\vec N}\,\,, \label{eq:4.18-a}
\ee
\be
{\cal S}^\Phi _i=-\frac 12\sum _{\vec M}M_i:\left( a_{\vec M}+a^{\dag }_
{-\vec M}\right) \left( a_{-\vec M}-a^{\dag }_{\vec M}\right) :\,\,{\rm for}
\,\,\eta |_{\partial \Sigma }=t_i\,\,.
\label{eq:4.18}
\ee
where $\vec N\times \vec M=(N_1M_2-N_2M_1)$.[There are similar expresions for
$S^2$, but they are long and will not be displayed here.]
These are also the generators one obtains from $l(\eta )$ with the choices
(\ref{eq:4.13})-(\ref{eq:4.14}) for $\eta $ if ${\cal Q}$ and ${\cal P}$ are
ignored.

It should not be a matter for surprise to note that $SDiff_0(\partial
\Sigma )$ does not mix the zero modes $a_0$ and $a_0^{\dag }$ with the rest,
and indeed that they are entirely absent, in (\ref{eq:4.18-a}) and
(\ref{eq:4.18}). For if $D$ is in this group with the action
$p\rightarrow Dp$ on points of $\partial \Sigma $, its action
$D^{*}:f\rightarrow D^{*}f$ on functions $f$ is by pull back:
$(D^{*}f)(p)=f(D(p))$. Constant functions being invariant under this action,
${\cal S}_{\vec N}^\Phi $ and ${\cal S}_i^\Phi $ must commute with $a_0$ and
$a_0^{\dag }$. It is for this reason that $a_0$ and $a_0^{\dag }$ do not occur
in ${\cal S}_{\vec N,i}^\Phi $.

We have now established the connection of our edge states to the modes of the
field $\Phi $, and at the same time motivated the dispersion relation $\omega
(n)=\omega _0$. But point b) is not yet fully covered, as we have not said
anything about the implementability of (\ref{eq:4.18-a}) and (\ref{eq:4.18})
on our Fock space. For this purpose, let us first consider the expression
(\ref{eq:4.18-a}) for ${\cal S}^\Phi _{\vec N}$. In (\ref{eq:4.18-a}), the
term with two creation and two annihilation operators commute with any
$a_{\vec L}$ and $a_{\vec L}^{\dag }$ for fixed $\vec L$ if $k$ is large
enough, suggesting that we can discard them. We can get zero for these term
also by noting that they (formally) change sign if we do the substitution
$\vec M=-\vec M'-\vec N$. For these reasons, we will discard the terms with
two creation and two annihilation operators in (\ref{eq:4.18-a}) and define
${\cal S}_{\vec N}^\Phi $ to be
\be
{\cal S}_{\vec N}^\Phi =\sum _{\vec M}\frac i2(\vec N\times \vec M)\left[
a_{-\vec M-\vec N}^{\dag }a_{-\vec M}-a_{\vec M+\vec N}a^{\dag }_{\vec M}
\right]\,.\label{eq:4.23}
\ee

As for ${\cal S}_i^\Phi $, if we regulate the sum as before by first summing
over $|M_i|\leq k$ and then letting $k\rightarrow \infty $, then the terms
with two creation and two annihilation operators vanish. We thus set
\be
{\cal S}_i^\Phi =\sum _{\vec M}M_ia_{\vec M}^{\dag }a_{\vec M}\,\,.
\label{eq:4.24}
\ee

Now we note that that the operators ${\cal S}_{\vec N}^\Phi $ and
${\cal S}_i^\Phi $ are well defined on the vacuum as they just annihilate the
latter. It is furthermore easy to see that they are well defined on any vector
in the Fock space.

We must also check the commutators and make sure that we do not get divergent
central terms. The calculation using (\ref{eq:4.23}) and (\ref{eq:4.24}) is
straightforward. There are no central terms, divergent or otherwise. We find
$$
\left[ S^\Phi _{\vec M},S^\Phi _{\vec N}\right] =i(\vec M\times \vec N)
S^\Phi _{\vec M+\vec N}\,,
$$
$$
\left[ S^\Phi _{\vec M},S^\Phi _i\right]=M_iS^\Phi _{\vec M}\,,
$$
\be
\left[ S^\Phi _i,S^\Phi _j\right] =0\,. \label{eq:4.27}
\ee

The Lagrangian (\ref{eq:1.8}) is not of course the only one
with $SDiff_0(\partial \Sigma )$ invariance, as we  can for instance replace
$\omega _0\varphi ^2$ by any potential $V(\varphi )$. We can in
particular change $\varphi ^2$ to $(\varphi -\varphi _0)^2$ for a constant
$\varphi _0$. This
model is of particular interest if the value of $A$ is macroscopic since the
shift
$\varphi \rightarrow \varphi +\varphi _0$  is not allowed as $A\rightarrow
\infty $ as in Lagrangians which spontaneously break symmetries. In any case,
for $A$ finite, (\ref{eq:1.8}) seems the simplest choice which accounts for
the properties of the edge states.

For the Lagrangian (\ref{eq:1.6}), $SDiff_0(\partial \Sigma )$ generators are
${\cal S}^\Phi _{\vec N}-iN_2Q_{\vec N}{\cal P}$ and ${\cal S}^\Phi _i$. It
is clear that this group continues to be implementable even after this
modification of ${\cal S}^\Phi _{\vec N}$.

It may be remarked that the $SDiff_0(\partial \Sigma )$ generators from
(\ref{eq:1.6}) and
(\ref{eq:1.7}) can be regularized if $\omega (n)$ approaches $\omega _0$ fast
enough as $J,\left| m\right| $ or $\left| N_i\right| $ become large. It is not
necessary that $\omega (n)$ is exactly a constant for all $n$. Similar
uncertainties exist for the dispersion relation of Chern-Simons edge
excitations. They have to be resolved using appropriate physical inputs.

\newpage
\sxn{THE ALGEBRA $w_{1+\infty }$}\label{sec-walgebra}

In this Section, we briefly consider the case $\Sigma =$ solid cylinder. it has
a cylinder $S^1\times \Re ^1$ as its boundary $\partial \Sigma $. Let $\theta $
(mod $2\pi $) and $z$ ($-\infty <z<\infty $) be coordinates on this $\partial
\Sigma $ and let
\be
\mu =d\theta dz\label{eq:5.1}
\ee
It is then known\cite{pope} that the Lie algebra of $SDff_0(\partial \Sigma )$
is the algebra $w_{1+\infty }$.

There is an easy way to see this result. Following earlier work \cite{pope},
we can regard $\mu $ as a symplectic form with the associated PB
\be
\left\{ f,g\right\} _\mu =\frac {\partial f}{\partial \theta }\frac
{\partial g}{\partial z}-\frac {\partial f}{\partial z}\frac {\partial g}
{\partial \theta }\,.\label{eq:5.2}
\ee
(This PB is different from the PB for (\ref{eq:1.6}), (\ref{eq:1.7}) or
(\ref{eq:1.8}).) The group $SDiff(\partial \Sigma )$ is then the group of
canonical transformations for the form $\mu $, the elements of its Lie algebra
being functions $f$ with Lie brackets given by (\ref{eq:5.2}). A basis for
this Lie algebra is the set of functions
\be
U_m^l=-ie^{im\theta }z^{l+1} \label{eq:5.3}
\ee
with the PB's
\be
\left\{ U_m^l,U_n^k\right\} _\mu = \left[ m(k+1)-n(l+1)\right] U_{m+n}^{l+k}
\label{eq:5.4}
\ee
which are exactly the defining relations of $w_{1+\infty }$.

We must now pass to quantum theory. For this purpose, we can either work with
the $\varphi $ Lagrangian (\ref{eq:1.8}) or equivalently use one of the two
Lagrangians (\ref{eq:1.6}) or (\ref{eq:1.7}) for $A$ and $B$. Let us work with
$\varphi $. Quantization requires a choice of basis $e_m$ for the Hilbert space
 ${\cal H}=L^2(\mu ,\partial \Sigma )$ of functions on $\partial \Sigma $ with
the scalar product
\be
\left( \alpha ,\beta \right) =\int _{\partial \Sigma }d\theta dz\alpha ^{*}
\beta (\theta ,z)\label{eq:5.5}
\ee
But identification of $e_m$ with the functions $U_m^l$ is not correct. They
are not normalizable if $l\geq -1/2$, nor do they have an interpretation along
the lines of plane waves in quantum mechanics. In this way, we are led to treat
$w_{1+\infty }$ in a basis different from (\ref{eq:5.3}). A simple basis
adopted to ${\cal H}$ is given by the correspondence
\be
e_n\rightarrow e_{n,N}=\frac {1}{\sqrt{2^NN!\pi ^{1/2}}}\exp (in\theta )
\exp (-\frac {z^2}2)H_N(z)\,\,\,n\in {\bf Z},\,\,N\in {\bf Z}^+,\label{eq:5.6}
\ee
$H_N$ being Hermite polynomials, and ${\bf Z}$ and ${\bf Z}^{+}$ the set of
integers
and of nonnegative integers respectively. $e_{n,N}$ has the symmetry
\be
e_{n,N}^{*}=e_{-n,N}.\label{eq:5.7}
\ee

We next expand the quantum versions $\Phi $ and $\dot \Phi $ of $\varphi $ and
$\dot \varphi $:
$$\Phi =\frac 1{\sqrt {2\omega _0}}\sum _{n,N}\left(
a_{n,N}e_{n,N}+a^{\dag }_{n,N}e_{n,N}^{*}\right) ,$$
\be
\dot \Phi =-i\sqrt {\frac {\omega _0}{2}}\sum _{n,N}\left(
a_{n,N}e_{n,N}-a^{\dag }_{n,N}e_{n,N}^{*}\right) .
\label{eq:5.8}\ee
Here the only nonzero commutator involving $a$'s and $a^{\dag }$'s is
\be
\left[ a_{n,N},a_{m,M}^{\dag }\right] =\delta _{nm}\delta _{NM}.
\label{eq:5.9}
\ee

The fields $\Phi $, $\dot \Phi $ are next realized by introducing a Fock space.
Let $\left| 0\right. \left\rangle {}\right. $ denote its vacuum.

The classical
generators of $SDff_0(\partial \Sigma )$ can be chosen to be
\begin{eqnarray}
s_{n,N} & = & \int d^3x\dot \varphi \left\{ e_{n,N},\varphi \right\} _\mu
\nonumber \\ & = & \int d^3x\dot \varphi \left( \frac {\partial e_{n,N}}
{\partial \theta }\frac \partial {\partial z}-\frac {\partial e_{n,N}}
{\partial z}\frac \partial {\partial \theta }\right) \varphi .
\label{eq:5.10}
\end{eqnarray}

On quantization they become
\be
S_{n,N} = :\int d^3x\dot \Phi \left( \frac {\partial e_{n,N}}{\partial
\theta
}\frac \partial {\partial z}-\frac {\partial e_{n,N}}{\partial z}\frac \partial
{\partial \theta }\right) \Phi : \label{eq:5.11}
\ee

Although it is of interest to compute the commutators of $S_{n,N}$, we will not
attempt that task here.
\newpage
\sxn{WHAT IS 3d HALL EFFECT}\label{sec-hall}

In Section 1, the relevance of discovering a three dimensional analogue of Hall
effect in the context of our problem was motivated. In this last Section, we
make a proposal for the same.

Let us briefly recall the Hall effect. Here there is a particle of charge $e$
in the $1-2$ plane in the presence of fields $F_{12}$ and $F_{a0}$ ($a=0,1$)
which we take to be time independent constants for simplicity. The particle is
subject to the force
\be
e\left( F_{a0}+\epsilon _{ab}\dot x^bF_{12}\right) ,\,\,\,\,\,\epsilon
_{ab}=\epsilon ^{0ab3}
\label{eq:6.1}
\ee
where $ x^a$ are its (Cartesian) coordinates and $\dot x^a$ the components of
its velocity. The force vanishes if
\be
\dot x^a=\epsilon ^{ab}\frac {F_{b0}}{F_{12}}\,\,,\,\,\,\, \epsilon ^{ab}
\equiv \epsilon _{ab},\label{eq:6.2}
\ee
and this equation embodies the Hall effect.

In three dimensions, what corresponds to $F_{12}$ is $H_{123}$. As it is the
vortex or the string which has coupling to this field, we look for the analogue
of (\ref{eq:6.2}) when vortices interact with $H$, or rather with $B$. This
interaction is \cite{ramond}
\be
{\cal L}_{INT}=\frac \lambda 2\epsilon ^{ab}B_{\mu \nu }(y)\partial _ay^\mu
(\sigma )\partial _by^\nu (\sigma ),\,\,\,\partial _a\equiv \frac {\partial }
{\partial \sigma ^a} \label{eq:6.3}
\ee
where $\sigma =(\sigma ^0,\sigma ^1),\,\sigma ^0$ is the evolution parameter,
$y:\sigma \rightarrow y(\sigma )$ describes the spacetime history of the
vortex, $\lambda $ is a constant and the Levi-Civita symbol is defined by
$\epsilon ^{01}=-\epsilon ^{10}=1$. [It is thus different from the
$\epsilon $'~s in (\ref{eq:6.1}) and (\ref{eq:6.2}).]

The ``force'' term from (\ref{eq:6.3}) and what corresponds to (\ref{eq:6.1})
is
\be
\frac {\partial {\cal L}_{INT}}{\partial y^\mu }-
\partial _a\frac {\partial {\cal L}_{INT}}{\partial (\partial _ay^\mu )}=
\frac \lambda 2H_{\mu \nu \lambda }(y)\epsilon ^{ab}
\partial _ay^\nu \partial _by^\lambda \label{eq:6.4}
\ee

which, with the choice $y^0(\sigma )=\sigma ^0$ becomes, for $\mu =i=1,2,3,$
\be
\lambda \left[ H_{123}\epsilon _{ijk}\partial _0y^j\partial _1y^k-H_{ij0}
\partial _1y^j\right],\,\,\,\,\,\,\,\,\epsilon _{ijk}\equiv \epsilon ^{oijk}.
\label{eq:6.5}
\ee
$H_{123}$, $H_{ij0}$ are here regarded as time independent constants as in
(\ref{eq:6.1}).

(\ref{eq:6.5}) vanishes for
\be
\left( \partial _0- c\partial _1\right) y^j=-\frac 12\epsilon ^{jkl}\frac
{H_{kl0}}{H_{123}}\,\, ,\,\,\,\,\,\, \epsilon^{jkl}=\epsilon ^{0jkl},
\label{eq:6.6}
\ee
$c$ being any real constant. It is this equation which describes the vortex
Hall effect. It implies the wave equation
\be
\left( \partial _0+c\partial _1\right) \left( \partial _0-c\partial
_1\right) y^j=0\label{eq:6.7}
\ee
and can thus provide solutions of the field equation for the Lagrangian
$$\int d^2\sigma \left( {\cal L}_0+{\cal L}_{INT}\right) ,$$
\be
{\cal L}_0=\frac 12\left[ \frac 1{c^2}\left( \partial _0y^j\right) ^2-\left(
\partial _1y^j\right) ^2\right] .\label{eq:6.8}
\ee

It seems reasonable to propose (\ref{eq:6.6}) as describing the three
dimensional analogue of the Hall effect. It displays the response of a vortex
to charge density $J^0=-2H_{123}$ and current density $J^j=\epsilon
^{jkl}H_{kl0}.$ Depending on the sign of $c$, it claims that the left or right
moving vortex mode has velocity parallel to current density.

It is easy to solve (\ref{eq:6.6}). The general solution is
\be
y^j(\sigma ^0,\sigma ^1)=-\frac 12\epsilon ^{jkl}\frac {H_{kl0}}{H_{123}}
(\sigma ^0-\frac 1c\sigma ^1)+z^j(\sigma ^0+\frac 1c\sigma ^1) \label{eq:6.9}
\ee
where the functions $z^j$ are not determined by (\ref{eq:6.6}) alone.

Now a closed vortex is periodic in $\sigma ^1$. In this case, then,
\be
z^j(\sigma ^0+\frac 1c\sigma ^1)=-\frac 12\epsilon ^{jkl}\frac {H_{kl0}}
{H_{123}}(\sigma ^0+\frac 1c\sigma ^1)+\tilde z^j(\sigma ^0+
\frac 1c\sigma ^1)\,\, ,\label{eq:6.10}
\ee
$\tilde z^j$ being a periodic function, and
\be
y^j(\sigma ^0,\sigma ^1)=-\epsilon ^{jkl}\frac {H_{kl0}}{H_{123}}\sigma
^0+\tilde z^j(\sigma ^0+\frac 1c\sigma ^1).\label{eq:6.11}
\ee

But if the vortex is not closed, and its ends terminates on $\partial \Sigma $,
we can not conclude (\ref{eq:6.10}) and (\ref{eq:6.11}).

\newpage
{\bf Acknowledgements}

We thank Ted Allen, Peppe Bimonte, Mark Bowick, Elisa Ercolessi, Kumar Gupta,
Amitabha Lahiri and Ajit Srivastava for discussions. This work was supported
by the Department of Energy under contract Number DE-FG02-85ER40231. P.T.S.
also thanks CAPES (Brazil) for partial support.

\end{document}